\documentclass[journal,twocolumn]{IEEEtran}
\usepackage{cmap}
\usepackage[T1]{fontenc}
\usepackage{amsmath,graphicx,hhline}
\usepackage{caption}
\usepackage{subcaption}
\usepackage{amssymb,amsthm}
\usepackage{mathrsfs}
\usepackage[nodisplayskipstretch]{setspace}
\usepackage{enumerate}
\usepackage{multirow}
\usepackage[update,prepend]{epstopdf}
\usepackage{float}
\restylefloat{table}
\usepackage{url}
\usepackage{listings}
\usepackage{algorithm}
\usepackage{algpseudocode}
\usepackage{booktabs}
\usepackage{lipsum}
\usepackage{geometry}

\newgeometry{
    top=1in,
    bottom=1in,
    outer=1in,
    inner=1in,
}

\newtheorem*{theorem*}{Theorem}
\newtheorem{theorem}{Theorem}
\newtheorem*{problemstatement*}{Problem Statement}
\newtheorem*{definition*}{Definition}
\newtheorem{definition}[theorem]{Definition}
\newtheorem{Lemma}[theorem]{Lemma}
\newtheorem*{Lemma*}{Lemma}
\newtheorem*{corollary*}{Corollary}

\newtheorem{property}[theorem]{Property}

\title{Graph Equivalence Classes for Spectral Projector-Based Graph {F}ourier Transforms}
\author{Joya A. Deri,~\IEEEmembership{Member, IEEE,}
        and Jos\'{e} M. F. Moura,~\IEEEmembership{Fellow, IEEE}
        \thanks{This work was partially supported by NSF grants CCF-1011903 and CCF-1513936 and an SYS-CMU grant.}
        \thanks{The authors are with the Department of Electrical and Computer Engineering, Carnegie Mellon University, Pittsburgh, PA 15213 USA (email: {jderi,moura}@andrew.cmu.edu)}%
        }

\begin{document}
\maketitle
\begin{abstract}
We define and discuss the utility of two equivalence graph classes over which a spectral projector-based graph Fourier transform is equivalent:  isomorphic equivalence classes and Jordan equivalence classes. Isomorphic equivalence classes show that the transform is equivalent up to a permutation on the node labels. Jordan equivalence classes  permit identical transforms over graphs of nonidentical topologies and allow a basis-invariant characterization of total variation orderings of the spectral components. Methods to exploit these classes to reduce computation time of the transform as well as limitations are discussed.
\end{abstract}
\begin{IEEEkeywords}
Jordan decomposition, generalized eigenspaces, directed graphs, graph equivalence classes, graph isomorphism, signal processing on graphs, networks
\end{IEEEkeywords}
\section{Introduction}
\label{sec:intro}
Graph signal processing~\cite{sandryhaila2013discrete,shuman2013emerging} permits applications of digital signal processing concepts  to increasingly larger networks. It is based on defining a shift filter, for example, the adjacency matrix in~\cite{sandryhaila2013discrete,sandryhaila2014big,sandryhaila2014discrete} to analyze undirected and directed graphs, or the graph Laplacian~\cite{shuman2013emerging} that applies to undirected graph structures. The graph Fourier transform is defined through the eigendecomposition of this shift operator, see these references. Further developments have been considered in~\cite{teke2017extending,zhu2012approximating,narang2012perfect}. In particular, filter design~\cite{sandryhaila2013discrete,teke2017extending,teke2017extending2} and sampling~\cite{marques2016sampling,segarra2016reconstruction,chen2015signal} can be applied to reduce the computational complexity of graph Fourier transforms.

With the objective of simplifying graph Fourier transforms for large network applications, this paper explores methods based on graph equivalence classes to reduce the computation time of the subspace projector-based graph Fourier transform proposed in~\cite{deriGFT2016}. This transform extends the graph signal processing framework proposed by~\cite{sandryhaila2013discrete,sandryhaila2014big,sandryhaila2014discrete} to consider spectral analysis over directed graphs with potentially non-diagonalizable (\emph{defective}) adjacency matrices. 
The graph signal processing framework of~\cite{deriGFT2016}  allows for a  unique, unambiguous signal representation over defective adjacency matrices.

Consider a graph $\mathcal G=G(A)$ with adjacency matrix $A\in\mathbb C^{N\times N}$ with $k\leq N$ distinct eigenvalues and Jordan decomposition $A=VJV^{-1}$.  The associated Jordan subspaces of $A$ are $\mathscr J_{ij}$, $i=1,\dots k$, $j=1,\dots,g_i$, where~$g_i$ is the geometric multiplicity of eigenvalue~$\lambda_i$, or the dimension of the kernel of $A-\lambda_i I$. The signal space $\mathcal S$ can be uniquely decomposed by the Jordan subspaces (see~\cite{lancaster1985,horn2012matrix} and Section~\ref{sec:background}).  For a graph signal $s\in\mathcal S$, the graph Fourier transform (GFT) of~\cite{deriGFT2016} is defined as
\begin{align}
\mathcal F: \mathcal S &\rightarrow  \bigoplus_{i=1}^k \bigoplus_{j=1}^{g_i} \mathscr J_{ij}\nonumber\\
s&\rightarrow \left(\widehat s_{11},\dots,\widehat s_{1g_1},\dots,\widehat s_{k1},\dots,\widehat s_{kg_k}\right) \label{eq:gft},
\end{align}
where $s_{ij}$ is the (oblique) projection of $s$ onto the Jordan subspace $J_{ij}$ parallel to $\mathcal S\backslash \mathscr J_{ij}$. That is, the Fourier transform of~$s$, is the unique decomposition
\begin{equation}
\label{eq:gft_sum}
s = \sum_{i=1}^k\sum_{j=1}^{g_i} \widehat s_{ij},\hspace{1cm}\widehat s_{ij}\in \mathscr J_{ij}.
\end{equation} The spectral components are the Jordan subspaces of the adjacency matrix with this formulation.

This paper presents graph equivalence classes where equal GFT projections by~\eqref{eq:gft} are the \emph{equivalence relation}. First, the transform~\eqref{eq:gft} is invariant to node permutations, which we formalize with the concept of isomorphic equivalence classes.
Furthermore, the GFT permits degrees of freedom in graph topologies, which we formalize by defining Jordan equivalence classes, a concept that allows graph Fourier transform computations over graphs of simpler topologies. A frequency-like ordering based on total variation  of the spectral components  is also presented to motivate low-pass, high-pass, and pass-band graph signals.

Section~\ref{sec:background} provides the graph signal processing and linear algebra background for the graph Fourier transform~\eqref{eq:gft}.  Isomorphic equivalence classes are defined in Section~\ref{sec:gft:isomorphic}, and Jordan equivalence classes are defined in Section~\ref{sec:jordanequiv}. The Jordan equivalence classes influence the definition of total variation-based orderings of the Jordan subspaces, which is discussed in detail in Section~\ref{sec:jordanequiv:totalvar}. Section~\ref{sec:example} illustrates Jordan equivalence classes and total variation orderings. Limitations of the method are discussed in Section~\ref{sec:limitations}.
%
%
%
%
\section{Background}
\label{sec:background}
This section reviews the concepts of graph signal processing and the GFT~\eqref{eq:gft}. Background on graphs signal processing, including definitions of graph signals and the graph shift, is described in greater detail in~\cite{sandryhaila2013discrete,sandryhaila2014big,sandryhaila2014discrete,deriGFT2016}. For background on eigendecompositions, the reader is directed to  in~\cite{lancaster1985,gohberg2006invariant,golub2013matrix}.
\subsection{Eigendecomposition}
\label{sec:background:eigendecomp}
Consider matrix~$A\in \mathbb C^{N\times N}$ with~$k$ distinct eigenvalues~$\lambda_1,\dots,\lambda_k$, $k\leq N$. 
The algebraic multiplicity~$a_i$ of~$\lambda_i$ represents the corresponding exponent of the characteristic polynomial of~$A$. Denote by $\mathrm{Ker}(A)$ the kernel or null space of matrix $A$. The \emph{geometric multiplicity}~$g_i$ of eigenvalue $\lambda_i$ equals the dimension of   $\mathrm{Ker} \left(A-\lambda_i I\right)$, which is the \emph{eigenspace} of $\lambda_i$ where $I$ is the $N\times N$ identity matrix.  The generalized eigenspaces $\mathscr G_i$, $i=1,\dots,k$, of $A$ are defined as
\begin{equation}
\label{eq:generalizedeigenspace}
\mathscr G_i = \mathrm{Ker}(A-\lambda_i I)^{m_i},
\end{equation}
where~$m_i$ is the index of eigenvalue~$\lambda_i$. The generalized eigenspaces uniquely decompose $\mathbb C^N$ as the direct sum
\begin{equation}
\label{eq:genspaces_decomp}
\mathbb C^N = \bigoplus_{i=1}^{k} \mathscr G_i.
\end{equation}

\textbf{Jordan chains.} Let $v_{1} \in \mathrm{Ker}(A - \lambda_i I)$, $v_1\neq 0$, be a proper eigenvector of~$A$ that generates generalized eigenvectors by the recursion
\begin{equation}
\label{eq:jordanchain}
Av_{p} = \lambda_i v_{p} + v_{p-1}, \:\: p=2,\dots,r
\end{equation}
where~$r$ is the minimal positive integer such that $\left(A-\lambda_i I\right)^r v_r=0$ and $\left(A-\lambda_i I\right)^{r-1} v_r\neq0$.
A sequence of vectors $(v_{1},\dots,v_{r})$  that satisfy~\eqref{eq:jordanchain} is a \emph{Jordan chain of length~$r$}~\cite{lancaster1985}. The vectors in a Jordan chain are linearly independent and generate the \emph{Jordan subspace}
\begin{equation}
\label{eq:jordansubspace}
\mathscr J = \mathrm{span}\left(v_1,v_2,\dots,v_r\right).
\end{equation}

Denote by $\mathscr J_{ij}$ the $j$th Jordan subspace of $\lambda_i$ with dimension $r_{ij}$, $i=1,\dots,k$, $j=1,\dots,g_i$. The Jordan spaces  are disjoint and uniquely decompose the generalized eigenspace~$\mathscr G_i$~\eqref{eq:generalizedeigenspace} of $\lambda_i$ as
\begin{equation}
\label{eq:jordansubspace_decomp}
\mathscr G_i = \bigoplus_{j = 1}^{g_i}\mathscr J_{ij}.
\end{equation}

The space~$\mathbb C^N$ can be expressed as the unique decomposition of Jordan spaces
\begin{equation}
\label{eq:CN-jordansubspace_decomp}
\mathbb C^N = \bigoplus_{i = 1}^k\bigoplus_{j = 1}^{g_i}\mathscr J_{ij}.
\end{equation}

\textbf{Jordan decomposition.} Let~$V_{ij}$ denote the $N\times r_{ij}$ matrix whose columns form a Jordan chain of eigenvalue~$\lambda_i$ that spans Jordan subspace~$\mathscr J_{ij}$.  Then the eigenvector matrix~$V$ of $A$ is
\begin{equation}
\label{eq:V}
V = \begin{bmatrix}
V_{11}  \cdots  V_{1g_1} & \cdots &V_{k1} \cdots V_{kg_k}
\end{bmatrix},
\end{equation}
where $k$ is the number of distinct eigenvalues. The columns of $V$ are a \emph{Jordan basis}  of $\mathbb C^N$. Then~$A$ has block-diagonal \emph{Jordan normal form}~$J$ consisting of Jordan blocks
\begin{equation}
\label{eq:jordanblock}
\setlength\arraycolsep{12pt}
J(\lambda) = \begin{bmatrix}
\lambda&1& \\
&\lambda&\ddots     \\
&&\ddots &1\\
&  && \lambda
\end{bmatrix}.
\end{equation}
of size~$r_{ij}$.  The Jordan normal form $J$ of $A$ is unique up to a permutation of the Jordan blocks. The \emph{Jordan decomposition} of~$A$ is $A = VJV^{-1}$. 
\subsection{Spectral Components}
\label{sec:background:spectralcomponents}
The spectral components of the Fourier transform~\eqref{eq:gft} are expressed in terms of the eigenvector basis $v_1,\dots,v_N$ and its dual basis $w_1,\dots,w_N$ since the Jordan basis may not be orthogonal. Denote the basis and dual basis matrices by $V = [v_1 \cdots v_N]$ and $W = [w_1 \cdots, w_N]$.  The dual basis matrix is the inverse Hermitian $W = V^{-H}$~\cite{horn2012matrix,jelena2014foundations}.

Consider the $j$th spectral component of $\lambda_i$
\begin{equation}
\mathscr J_{ij} = \mathrm{span}(v_1,\cdots v_{r_{ij}}).
\end{equation}
The \emph{projection matrix}  onto $\mathscr J_{ij}$ parallel to $\mathbb C^N\backslash\mathscr J_{ij}$ is
\begin{equation}
\label{eq:spectralprojectormatrix}
P_{ij} = V_{ij} W^H_{ij},
\end{equation}
where
\begin{equation} V_{ij}= [v_1 \cdots v_{r_{ij}}]\end{equation} is the corresponding submatrix of $V$ and $W^H_{ij}\in\mathbb C^{r_{ij}\times N}$ is the corresponding submatrix of $W$ partitioned as
\begin{equation}W = [\cdots W_{i1}^H\cdots  W_{ig_i}^H \cdots]^T.\end{equation}

As shown in~\cite{deriGFT2016}, the projection of signal $s\in\mathbb C^N$ onto Jordan subspace~$\mathscr J_{ij}$ can be written as
\begin{align}
\widehat s_{ij}  &= \widetilde s_1 v_1 + \dots +\widetilde s_{r_{ij}} v_{r_{ij}} \label{eq:gft_singlecomponent_1}\\
&= V_{ij} W^H_{ij} s.
\label{eq:gft_singlecomponent}
\end{align}

The next sections show that invariance of the graph Fourier transform~\eqref{eq:gft} is a useful \emph{equivalence relation} on a set of graphs. Equivalence classes with respect to the GFT are explored in Sections~\ref{sec:gft:isomorphic} and~\ref{sec:jordanequiv}.
\section{Isomorphic Equivalence Classes}
\label{sec:gft:isomorphic}
This section demonstrates that the graph Fourier transform~\eqref{eq:gft} is invariant up to a permutation of node labels and establishes sets of isomorphic graphs as equivalence classes with respect to invariance of the GFT~\eqref{eq:gft}. Two graphs $\mathcal G(A)$ and $\mathcal G(B)$ are isomorphic if their adjacency matrices are similar with respect to a permutation matrix~$T$, or $B = TAT^{-1}$~\cite{cvetkovic1988recent}. The graphs have the same Jordan normal form and the same spectra. Also, if $V_A$ and~$V_B$ are eigenvector matrices of~$A$ and~$B$, respectively, then $V_B = TV_A$. We prove that the set~$\mathbf G_A^I$ of all graphs that are isomorphic to $\mathcal G(A)$ is an equivalence class over which the GFT is preserved. The next theorem shows that an appropriate permutation can be imposed on the graph signal and GFT to ensure invariance of the GFT over all graphs $\mathcal G\in \mathbf G_A^I$.
\begin{theorem}
\label{thm:isomorphicinvariance}
The graph Fourier transform of a signal $s$ is invariant to the choice of graph $\mathcal G\in\mathbf G_A^I$ up to a permutation on the graph signal and inverse permutation on the graph Fourier transform.
\end{theorem}
\begin{IEEEproof}
For $\mathcal G(A),\mathcal G(B)\in\mathbf G_A^I$, there exists a permutation matrix $T$ such that $B = TAT^{-1}$. For eigenvector matrices $V_A$ and $V_B$ of $A$ and $B$, respectively, let $V_{A,ij}$ and~$V_{B,ij}$ denote the $N\times r_{ij}$ submatrices of~$V_A$ and~$V_B$ whose columns span the $j$th Jordan subspaces~$\mathscr J_{A,ij}$ and~$\mathscr J_{B,ij}$ of the $i$th eigenvalue of $A$ and $B$, respectively.  Let $W_A = V_A^{-H}$ and $W_B = V_B^{-H}$ denote the matrices whose columns form dual bases of $V_A$ and $V_B$. Since $V_B = T V_A$,
\begin{align}
W_B &= (TV_A)^{-H} \\
&= (V_A^{-1}T^{-1})^H\\
&= T^{-H} V_A^{-H}\\
&= T W_A,
\end{align}
where $T^{-H} = T$ since $T$ is a permutation matrix. Thus, \begin{equation}W_B^H = W_A^H T^H = W_A^H T^{-1}.\end{equation}

Consider graph signal $s$. By~\eqref{eq:gft_singlecomponent}, the signal projection onto~$\mathscr J_{A,ij}$ is
\begin{equation}
\label{eq:isomorphic_A}
\widehat s_{A,ij} = V_{A,ij} W_{A,ij}^Hs.
\end{equation}
Permit a permutation $\overline s = Ts$ on the graph signal. Then the projection of $\overline s$ onto $\mathscr J_{B,ij}$ is
\begin{align}
\widehat{\overline{s}}_{B,ij} &= T V_{A,ij} W_{A,ij}^H T^{-1} Ts\\
&= T V_{A,ij} W_{A,ij}^H s\\
&= T  \widehat s_{A,ij}
\end{align}
by~\eqref{eq:isomorphic_A}.
Therefore, the graph Fourier transform~\eqref{eq:gft} is invariant to a choice among isomorphic graphs up to a permutation on the graph signal and inverse permutation on the Fourier transform.
\end{IEEEproof}
\begin{theorem}
\label{thm:isomorphicinvariance2}
Consider $A\in\mathbb C^{N\times N}$. Then the set $\mathbf G_A^I$ of graphs isomorphic to $\mathcal G(A)$  is an equivalence class with respect to the invariance of the GFT~\eqref{eq:gft} up to a permutation of the graph signal and inverse permutation of the graph Fourier transform.
\end{theorem}
Theorem~\ref{thm:isomorphicinvariance} establishes an invariance of the GFT over graphs that only differ up to a node labeling, and
Theorem~\ref{thm:isomorphicinvariance2} follows.

The isomorphic equivalence of graphs is important since it signifies that the rows and columns of an adjacency matrix can be permuted to accelerate the eigendecomposition. For example, permutations of highly sparse adjacency matrices can convert an arbitrary matrix to nearly diagonal forms, such as with the Cuthill-McKee algorithm~\cite{Cuthill:1969:RBS:800195.805928}. Optimizations for such matrices in this form are discussed in~\cite{golub2013matrix} and~\cite{han2000comparison}, for example.   In the next section, the degrees of freedom in graph topology are explored to define another GFT equivalence class.
%
%
%
%
\section{Jordan Equivalence Classes}
\label{sec:jordanequiv}
Since the Jordan subspaces of defective adjacency matrices are nontrivial (i.e., they have dimension larger than one), a degree of freedom exists on the graph structure so that the graph Fourier transform of a signal is equal over multiple graphs of different topologies. This section defines \emph{Jordan equivalence classes} of graph structures over which the GFT~\eqref{eq:gft} is equal for a given graph signal.  The section proves important properties of this equivalence class that are used to explore inexact methods and real-world applications in~\cite{deriAIM2016}.

The intuition behind Jordan equivalence is presented in Section~\ref{sec:jordanequiv:intuition}, and properties of Jordan equivalence are described in Section~\ref{sec:jordanequiv:def}.  Section~\ref{sec:jordanequiv:Viso} compares isomorphic and Jordan equivalent graphs. Sections~\ref{sec:jordanequiv:diagonaljordanform},~\ref{sec:jordanequiv:onejordanblock},~\ref{sec:jordanequiv:twojordanblocks},  and~\ref{sec:jordanequiv:multipleblocks} prove properties for Jordan equivalence classes when adjacency matrices have particular Jordan block structures. 
\subsection{Intuition}
\label{sec:jordanequiv:intuition}
\begin{figure}[tb]
\centering
\includegraphics[width=.7\linewidth]{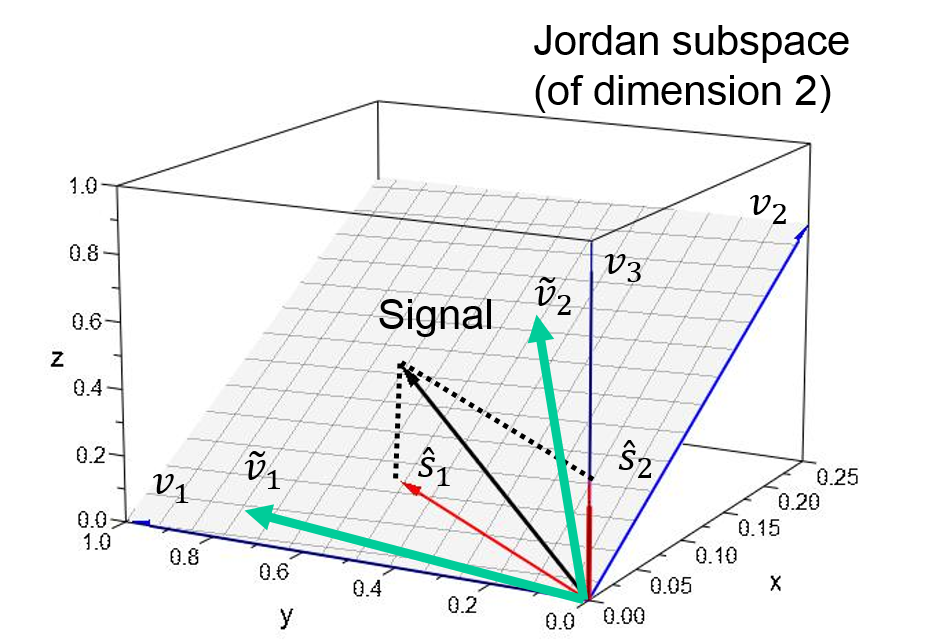}
\caption{\small Projections $\widehat s_1$ and $\widehat s_2$ (shown in red) of a signal $s$ (shown in black) onto a nontrivial Jordan subspace (span of $v_1$ and $v_2$) and the span of~$v_3$, respectively, in $\mathbb R^3$. The projection onto the nontrivial subspace is invariant to basis choices $\{v_1,v_2\}$ (in blue) or $\{\widetilde v_1,\widetilde{v}_2\}$ (in green).}
\label{fig:jordanequiv:intuition}
\end{figure}

Consider Figure~\ref{fig:jordanequiv:intuition}, which shows a basis $\{V\}=\{v_1,v_2,v_3\}$ of $\mathbb R^3$ such that $v_2$ and $v_3$ span a two-dimensional Jordan space $\mathscr J$ of adjacency matrix~$A$ with Jordan decomposition $A=VJV^{-1}$. The resulting projection of a signal $s\in\mathbb R^N$ as in~\eqref{eq:gft_singlecomponent} is unique.

Note that the definition of the two-dimensional Jordan subspace~$\mathscr J$ in Figure~\ref{fig:jordanequiv:intuition} is not basis-dependent because any spanning set $\{w_2,w_3\}$ could be chosen to define $\mathscr J$. This can be visualized by rotating $v_2$ and $v_3$ on the two-dimensional plane. Any choice  $\{w_2,w_3\}$ corresponds to a new basis $\widetilde V$. 
Note that $\widetilde A = \widetilde V J \widetilde V^{-1}$  does not equal $A=V JV^{-1}$ for all choices of $\{w_2,w_3\}$; the underlying graph topologies may be different, or the edge weights may be different. Nevertheless, their spectral components (the Jordan subspaces) are identical, and,  consequently, the spectral projections of a signal onto these components are identical; i.e., the GFT~\eqref{eq:gft} is equivalent over graphs $\mathcal G(A)$ and $\mathcal G(\widetilde A)$. This observation leads to the definition of \emph{Jordan equivalence classes} which preserve the GFT~\eqref{eq:gft} as well as the underlying structure captured by the Jordan normal form~$J$ of~$A$. These classes are formally defined in the next section.
\subsection{Definition and Properties}
\label{sec:jordanequiv:def}
This section defines the \emph{Jordan equivalence class} of graphs, over which the graph Fourier transform~\eqref{eq:gft} is invariant. We will show that certain Jordan equivalence classes allow the GFT computation to be simplified.

Consider graph~$\mathcal G(A)$ where $A$ has a Jordan chain that spans Jordan subspace~$\mathscr J_{ij}$ of dimension $r_{ij}> 1$. Then~\eqref{eq:gft_singlecomponent_1}, and consequently,~\eqref{eq:gft_singlecomponent}, would hold for a non-Jordan basis of $\mathscr J_{ij}$; that is, a basis could be chosen to find spectral component~$\widehat s_{ij}$ such that the basis vectors do not form a Jordan chain of $A$. This  highlights that the Fourier transform~\eqref{eq:gft} is characterized not by the Jordan basis of~$A$ but by the set~${\mathbf J}_A=\{\mathscr J_{ij}\}_{ij}$ of {Jordan subspaces} spanned by the Jordan chains of~$A$. Thus, graphs with topologies yielding the same Jordan subspace decomposition of the signal space have the same spectral components. Such graphs are termed \emph{Jordan equivalent} with the following formal definition.

\begin{definition}[Jordan Equivalent Graphs]
\label{def:equivgraph}
Consider graphs~$\mathcal G(A)$ and~$ \mathcal G(B)$ with adjacency matrices $A,B\in\mathbb C^{N\times N}$. Then $\mathcal G(A)$ and $\mathcal G(B)$ are \emph{Jordan equivalent graphs} if all of the following are true:
\begin{enumerate}
\item $\mathbf J_A = \mathbf J_B$; and
\item $J_A= J_B$ (with respect to a fixed permutation of Jordan blocks). 
\end{enumerate}
\end{definition}

Let $\mathbf G_A^J$ denote the set of graphs that are Jordan equivalent to~$\mathcal G(A)$. Definition~\ref{def:equivgraph} and~\eqref{eq:gft} establish that $\mathbf G_A^J$ is an equivalence class.
\begin{theorem}
\label{thm:jordanequivclass}
For $A\in\mathbb C^{N\times N}$, the set $\mathbf G_A^J$ of all graphs that are Jordan equivalent to $\mathcal G(A)$ is an equivalence class with respect to invariance of the GFT~\eqref{eq:gft}.
\end{theorem}

Jordan equivalent graphs have adjacency matrices with identical Jordan subspaces and identical Jordan normal forms. This implies equivalence of graph spectra, proven in Theorem~\ref{thm:cospectral} below.
\begin{theorem}
\label{thm:cospectral}
Denote by $\Lambda_A$ and $\Lambda_B$ the sets of eigenvalues of $A$ and $B$, respectively. Let $\mathcal G(A),\mathcal G(B) \in\mathbf G_A^J$.  Then $\Lambda_A = \Lambda_B$; that is, $\mathcal G(A)$ and $\mathcal G(B)$ are cospectral.
\end{theorem}
\begin{IEEEproof}
Since $\mathcal G(A)$ and $\mathcal G(B)$ are Jordan equivalent, their Jordan forms are equal, so their spectra (the unique elements on the diagonal of the Jordan form) are equal.
\end{IEEEproof}

Once a Jordan decomposition for an adjacency matrix is found, it is useful to characterize other graphs in the same Jordan equivalence class.  To this end, Theorem~\ref{thm:jordanequiv_transformation} presents a transformation that preserves the Jordan equivalence class of a graph.
\begin{theorem}
\label{thm:jordanequiv_transformation}
Consider $A,B\in\mathbb C^{N\times N}$ with Jordan decompositions $A = VJV^{-1}$ and $B = XJX^{-1}$ and eigenvector matrices $V =[V_{ij}]$ and $X=[X_{ij}]$, respectively. Then, $\mathcal G(B)\in\mathbf G_A^J$ if and only if $B$ has eigenvector matrix $X=VY$ for block diagonal $Y$ with invertible submatrices $Y_{ij}\in\mathbb C^{r_{ij}\times r_{ij}}$, $i=1,\dots,k$, $j=1,\dots,g_i$.
\end{theorem}
\begin{IEEEproof}
The Jordan normal forms of $A$ and $B$ are equal. By Definition~\ref{def:equivgraph}, it remains to show $\mathbf J_A = \mathbf J_B$ so that $\mathcal G(B)\in\mathbf G_A^J$.
The identity $\mathbf J_A = \mathbf J_B$ must be true when $\mathrm{span}\{V_{ij}\} = \mathrm{span}\{X_{ij}\} =\mathscr J_{ij}$, which implies that $X_{ij}$ represents an invertible linear transformation of the columns of $V_{ij}$. Thus, $X_{ij} = V_{ij}Y_{ij}$, where $Y_{ij}$ is invertible.  Defining $Y=\mathrm{diag}(Y_{11},\dots,Y_{ij},\dots,Y_{k,g_k})$ yields $X= VY$.
\end{IEEEproof}

\subsection{Jordan Equivalent Graphs vs. Isomorphic Graphs}
\label{sec:jordanequiv:Viso}
This section shows that isomorphic graphs do not imply Jordan equivalence, and vice versa. 
First it is shown that isomorphic graphs have isomorphic Jordan subspaces.
\begin{Lemma}
\label{lemma:jordanequiv_isomorphicsubspaces}
Consider graphs $\mathcal G(A),\mathcal G(B)\in\mathbf G_A^I$ so that $B = TAT^{-1}$ for a permutation matrix $T$. Denote by $\mathbf J_A$  and~$\mathbf J_B$ the sets of Jordan subspaces for $A$ and $B$, respectively. If $\{v_1,\dots,v_{r}\}$ is a basis of $\mathscr J_A\in\mathbf J_A$, then there exists $\mathscr J_B\in \mathbf J_B$ with basis $\{x_1,\dots,x_{r}\}$ such that $[x_1\cdots x_r] = T [v_1\cdots v_r]$; i.e., $A$ and $B$ have isomorphic Jordan subspaces.
\end{Lemma}
\begin{IEEEproof}
Consider $A$ with Jordan decomposition $A= VJV^{-1}$. Since $B = TAT^{-1}$, it follows that
\begin{align}
B & = T V JV^{-1} T^{-1}\\
&= XJX^{-1}\label{eq:isomorphiccheck}
\end{align}
where $X = TV$ represents an eigenvector matrix of $B$ that is a permutation of the rows of $V$. (It is clear that the Jordan forms of $A$ and $B$ are equivalent.) Let columns $v_1,\dots,v_r$ of~$V$ denote a Jordan chain of $A$ that spans Jordan subspace~$\mathscr J_A$. The corresponding columns in $X$ are $x_1,\dots,x_r$ and $\mathrm{span}(x_1,\dots,x_r)=\mathscr J_B$. Since $[x_1\ \cdots\ x_r] = T[v_1\ \cdots\  v_r]$,  $\mathscr J_A$ and $\mathscr J_B$ are isomorphic subspaces~\cite{lancaster1985}.
\end{IEEEproof}
\begin{theorem}
\label{thm:jordanequiv_isomorphic}
A graph isomorphism does not imply Jordan equivalence.
\end{theorem}
\begin{IEEEproof}
Consider $\mathcal G(A),\mathcal G(B) \in \mathbf G_A^I$ and $B= T A T^{-1}$ for permutation matrix $T$. By~\eqref{eq:isomorphiccheck}, $J_A = J_B$. To show $\mathcal G(A),\mathcal G(B)\in\mathbf G_A^J$, it remains to check whether $\mathbf J_A = \mathbf J_B$.

By Lemma~\ref{lemma:jordanequiv_isomorphicsubspaces}, for any $\mathscr J_A\in\mathbf J_A$, there exists $\mathscr J_B\in\mathbf J_B$ that is isomorphic to $\mathscr J_A$. That is, if $v_1,\dots,v_{r}$ and $x_1,\dots,x_{r}$ are bases of $\mathscr J_A$ and $\mathscr J_B$, respectively, then $[x_1\cdots x_r] = T [v_1\cdots v_r]$. Checking $\mathscr J_A = \mathscr J_B$ is equivalent to checking
\begin{align}
\alpha_1 v_1 + \dots + \alpha_r v_r &= \beta_1 x_1 + \dots + \beta_r x_r\\
&=\beta_1 Tv_1 + \dots + \beta_r Tv_r\label{eq:isomorphic_JAeqJB}
\end{align}
for some coefficients $\alpha_i$ and $\beta_i$, $i=1,\dots, r$. However,~\eqref{eq:isomorphic_JAeqJB} does not always hold. Consider  matrices $A$ and $B$
\begin{equation}
\label{eq:isomorphiccounterex}
\setlength\arraycolsep{10pt}
\renewcommand{\arraystretch}{.9}
A = \begin{bmatrix}
2 & 0 & -1\\ 0 & 2 & -1\\ 0 & 0 & 1
\end{bmatrix},
B = \begin{bmatrix}
1  &   0 &   0\\ -1  &   2 &   0\\ -1  &   0  &  2
\end{bmatrix}.
\end{equation}
These matrices are similar with respect to a permutation matrix and thus correspond to isomorphic graphs. Their  Jordan normal forms are both
\begin{equation}
\label{eq:isomorphiccounterex_J}
\renewcommand{\arraystretch}{.9}
\setlength\arraycolsep{12pt}
J = \begin{bmatrix}
1 & 0 & 0\\ 0 & 2 & 0\\ 0 & 0 & 2
\end{bmatrix}
\end{equation}
with possible eigenvector matrices $V_A$ and $V_B$ given by
\begin{equation}
\label{eq:isomorphiccounterex_V}
\renewcommand{\arraystretch}{.8}
\setlength\arraycolsep{10pt}
V_A = \begin{bmatrix}
1 & 1 & 0\\ 1 & 0 & 1\\ 1 & 0 & 0
\end{bmatrix},
V_B = \begin{bmatrix}
1  &   0 &   0\\ 1  &   1 &   0\\ 1  &   0  &  1
\end{bmatrix}.
\end{equation}
Equation~\eqref{eq:isomorphiccounterex_V} shows that $A$ and $B$ both have Jordan subspaces $\mathscr J_1= \mathrm{span}([1\ 1\ 1]^T)$ for  $\lambda_1=1$ and $\mathscr J_{21} = \mathrm{span}([0\ 1\ 0]^T)$ for one Jordan subspace of $\lambda_2 =2$. However, the remaining Jordan subspace is  $\mathrm{span}([1\ 0\ 0]^T)$ for $A$ but  $\mathrm{span}([0\ 0\ 1]^T)$ for $B$, so~\eqref{eq:isomorphic_JAeqJB} fails. Thus,~$\mathcal G(A)$ and $\mathcal G(B)$ are not Jordan equivalent.
\end{IEEEproof}

\begin{figure}[tb]
\centering
\includegraphics[width=.5\linewidth]{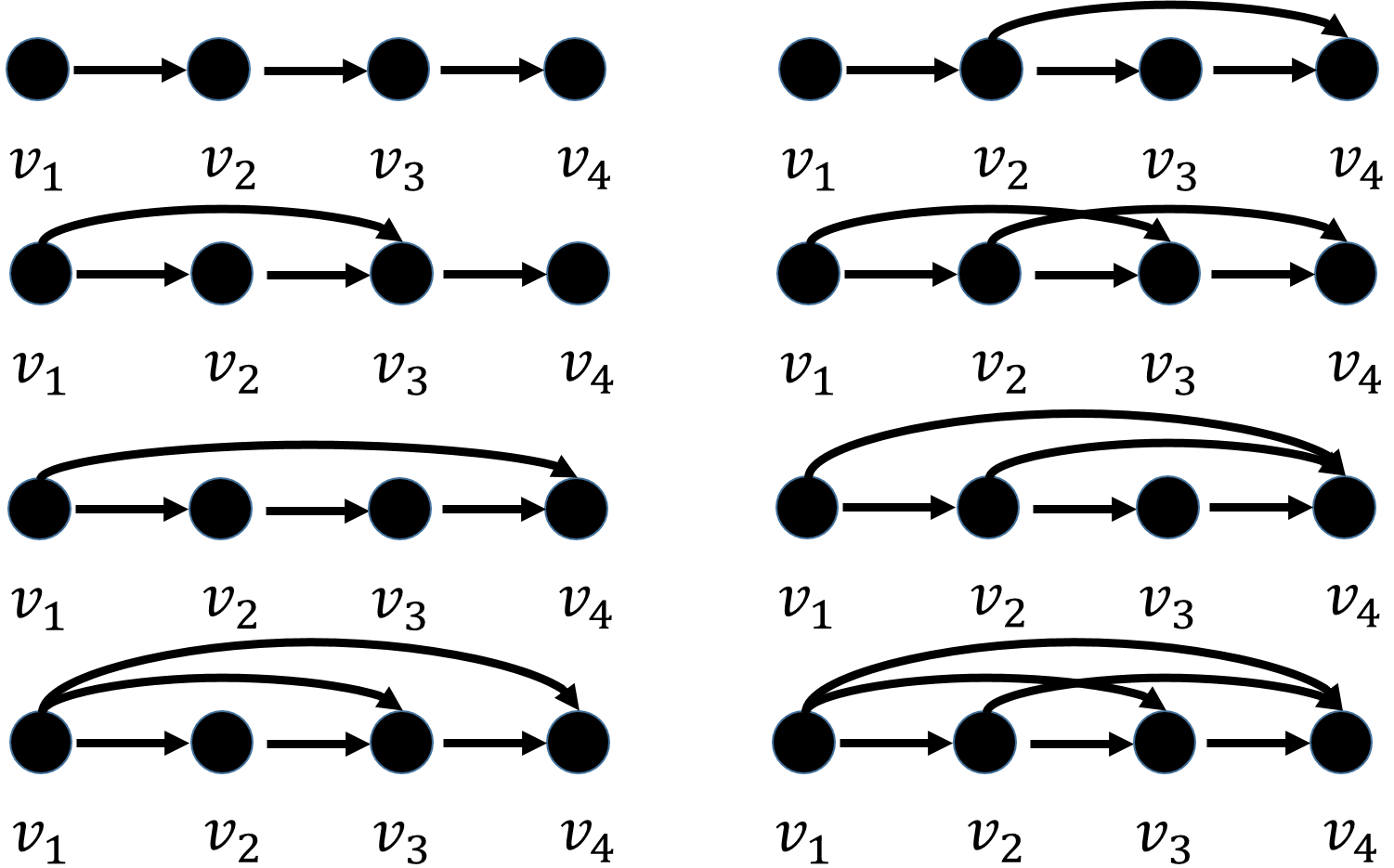}
\caption{\small Jordan equivalent graph structures with unicellular adjacency matrices.}
\label{fig:equivgraphs_unicell}
\vspace{-.10cm}
\end{figure}

The next theorem shows that Jordan equivalent graphs may not be isomorphic.
\begin{theorem}
\label{thm:jordanequiv_toisomorphic}
Jordan equivalence does not imply the existence of a graph isomorphism.
\end{theorem}
\begin{IEEEproof}
A counterexample is provided. The top two graphs in Figure~\ref{fig:equivgraphs_unicell} correspond to 0/1 adjacency matrices with a single Jordan subspace $\mathscr J = \mathbb C^N$ and eigenvalue~$0$; therefore, they are Jordan equivalent. On the other hand, they are not isomorphic since the graph on the right has more edges then the graph on the left.
\end{IEEEproof}

Theorem~\ref{thm:jordanequiv_isomorphic} shows that changing the graph node labels may change the Jordan subspaces and the Jordan equivalence class  of the graph, while Theorem~\ref{thm:jordanequiv_toisomorphic} shows that a Jordan equivalence class may include graphs with different topologies. Thus, graph isomorphism and Jordan equivalence are not identical concepts. Nevertheless, the isomorphic and Jordan equivalence classes both imply invariance of the graph Fourier transform with respect to equivalence relations as stated in Theorems~\ref{thm:isomorphicinvariance} and~\ref{thm:jordanequivclass}.

The next theorem establishes an isomorphism between Jordan equivalence classes.
\begin{theorem}
\label{thm:jordanequiv_isomorphism}
If $A,B\in\mathbb C^{N\times N}$ and $\mathcal G(A)$ and~$\mathcal G(B)$ are isomorphic, then their respective Jordan equivalence classes $\mathbf G^J_A$ and $\mathbf G^J_B$ are isomorphic; i.e., any graph~$\mathcal G(A')\in\mathbf G_A^J$ is isomorphic to a graph~$\mathcal G(B')\in\mathbf G_B^J$.
\end{theorem}
\begin{IEEEproof}
Let $\mathcal G(A)$ and $\mathcal G(B)$ be isomorphic by permutation matrix $T$ such that $B = TAT^{-1}$. Consider $\mathcal G(A')\in \mathbf G_A^J$, which implies that Jordan normal forms $J_{A'} = J_A$ and sets of Jordan subspaces $\mathbf J_{A'} =\mathbf J_A$ by Definition~\ref{def:equivgraph}. Denote by $A' = V_{A'}J_{A'}V_{A'}$ the Jordan decomposition of $A'$.  Define $B' = TA'T^{-1}$. It suffices to show $\mathcal G(B')\in\mathbf G_B^J$.  First simplify:
\begin{align}
B' &= TA'T^{-1}\\
&=TV_{A'} J_{A'} V_{A'}^{-1}T^{-1}\\
&= TV_{A'} J_{A} V_{A'}^{-1}T^{-1}\hspace{.5cm}\text{(since $\mathcal G(A')\in\mathbf G_A^J$)}\\
&= TV_{A'} J_{B} V_{A'}^{-1}T^{-1}\hspace{.5cm}\text{(since $\mathcal G(A)\in\mathbf G_B^I$)}.\label{eq:thm:jordanequiv_isomorphism}
\end{align}
From~\eqref{eq:thm:jordanequiv_isomorphism}, it follows that $J_{B'} = J_{B}$. It remains to show that $\mathbf J_{B'} = \mathbf J_B$. Choose arbitrary Jordan subspace $\mathscr J_{A,ij} = \mathrm{span}\{ V_{A,ij}\}$ of $A$. Then $\mathscr J_{A',ij} = \mathrm{span}\{ V_{A',ij}\} = \mathscr J_{A,ij}$ since $\mathcal G(A')\in\mathbf G_A^J$. Then the $j$th Jordan subspace of eigenvalue $\lambda_i$ for $B$ is
\begin{align}
\mathscr J_{B,ij} &= \mathrm{span}\{TV_{A,ij}\}\\
&= T \mathrm{span}\{V_{A,ij}\}.\label{eq:thm:jordanequiv_inter}
\end{align}
For the $j$th Jordan subspace of eigenvalue $\lambda_i$ for $B'$, it follows from~\eqref{eq:thm:jordanequiv_isomorphism} that
\begin{align}
\mathscr J_{B',ij} &= \mathrm{span}\{TV_{A',ij}\}\\
&= T \mathrm{span}\{V_{A',ij}\}\\
&= T \mathrm{span}\{V_{A,ij}\}\hspace{1cm}\text{(since $\mathcal G(A')\in\mathbf G_A^J$)}\\
&= \mathscr J_{B,ij}. \hspace{2cm}\text{(by~\eqref{eq:thm:jordanequiv_inter})}\label{eq:thm:jordanequiv_final}
\end{align}
Since~\eqref{eq:thm:jordanequiv_final} holds for all $i$ and $j$, the sets of Jordan subspaces $\mathbf J_{B'} = \mathbf J_B$. Therefore, $\mathcal G(B')$ and $\mathcal G(B)$ are Jordan equivalent, which proves the theorem.
\end{IEEEproof}

Theorem~\ref{thm:jordanequiv_isomorphism} shows that the Jordan equivalence classes of two isomorphic graphs are also isomorphic. This result permits an frequency ordering on the spectral components of a matrix~$A$ that is invariant to both the choice of graph in $\mathbf G_A^J$ and the choice of node labels, as demonstrated in Section~\ref{sec:jordanequiv:totalvar}.

\textbf{Relation to matrices with the same set of invariant subspaces.} Let $\mathbf G_A^{\mathrm{Inv}}$ denote the set of all matrices with the same set of invariant subspaces of $A$; i.e., $\mathcal G(B)\in\mathbf G_A^{\mathrm{Inv}}$ if and only if $\mathrm{Inv}(A) = \mathrm{Inv} (B)$. 
The next theorem shows that $\mathbf G_A^{\mathrm{Inv}}$  is a proper subset of the Jordan equivalence class $\mathbf G_A^J$ of $A$.
\begin{theorem}
\label{thm:jordanequiv_sameinvariantspaces}
For $A\in\mathbb C^{N\times N}$, $\mathbf G_A^{\mathrm{Inv}}\subset \mathbf G_A^J$.
\end{theorem}
\begin{IEEEproof}
If $\mathcal G(B)\in \mathbf G_A^{\mathrm{Inv}}$, then the set of Jordan subspaces are equal, or $\mathbf J_A=\mathbf J_B$.
\end{IEEEproof}

Theorem~\ref{thm:jordanequiv_sameinvariantspaces} sets the results of this chapter apart from analyses such as those in Chapter~10 of~\cite{gohberg2006invariant}, which describes structures for matrices with the same invariant spaces,  and~\cite{candan2011eigenstructure}, which describes the eigendecomposition of the discrete Fourier transform matrix in terms of projections onto invariant spaces. The Jordan equivalence class relaxes the assumption that \emph{all} invariant subspaces of two adjacency matrices must be equal. This translates to more degrees of freedom in the graph topology. The following sections present results for  adjacency matrices with diagonal Jordan forms, one Jordan block, and multiple Jordan blocks.
\subsection{Diagonalizable Matrices}
\label{sec:jordanequiv:diagonaljordanform}
If the canonical Jordan form~$J$ of~$A$ is diagonal ($A$ is diagonalizable), then there are no Jordan chains and the set of Jordan subspaces $\mathbf J_A = \{\mathscr J_p\}_{p=1}^N$ where $\mathscr J_p = \mathrm{span}(v_p)$ and $v_p$ is the $p$th eigenvector of~$A$. Graphs  with diagonalizable adjacency matrices include undirected graphs, directed cycles, and other digraphs with normal adjacency matrices such as normally regular digraphs~\cite{jorgensen1994normally}. A graph with a diagonalizable adjacency matrix is Jordan equivalent only to itself, as proven next.
\begin{theorem}
\label{thm:jordanequiv_diagonal}
A graph $\mathcal G(A)$ with diagonalizable adjacency matrix $A\in\mathbb C^{N\times N}$ belongs to a Jordan equivalence class of size one.
\end{theorem}
\begin{IEEEproof}
Since the Jordan subspaces  of a diagonalizable matrix are one-dimensional, the possible choices of Jordan basis are limited to nonzero scalar multiples of the eigenvectors. Then, given eigenvector matrix~$V$ of~$A$, all possible eigenvector matrices of $A$ are given by $X = VU$, where $U$ is a diagonal matrix with nonzero diagonal entries. Let $B = XJX^{-1}$, where $J$ is the diagonal canonical Jordan form of $A$. Since $U$ and $J$ are both diagonal, they commute, yielding
\begin{align}
B & = XJX^{-1}\\
&= VUJU^{-1}V^{-1}\\
&= VJUU^{-1}V^{-1}\\
&= VJV^{-1}\\
&= A.
\end{align}
Thus, a graph with a diagonalizable adjacency matrix is the one and only element in its Jordan equivalence class.
\end{IEEEproof}

When a matrix has nondefective but repeated eigenvalues, there are infinitely many choices of eigenvectors~\cite{golub2013matrix}. An illustrative example is the identity matrix, which has a single eigenvalue but is diagonalizable. Since it has infinitely many choices of eigenvectors, the identity matrix corresponds to infinitely many Jordan equivalence classes. By Theorem~\ref{thm:jordanequiv_diagonal}, each of these equivalence classes have size one. This observation highlights that the definition of a Jordan equivalence class requires a choice of basis.

\subsection{One Jordan Block}
\label{sec:jordanequiv:onejordanblock}
Consider matrix $A$ with Jordan decomposition $A = VJV^{-1}$ where $J$ is a single Jordan block and $V = [v_1\cdots v_N]$ is an eigenvector matrix. Then $A$ is a representation of a \emph{unicellular transformation} $T:\mathbb C^N\rightarrow \mathbb C^N$  with respect to Jordan basis $v_1,\dots v_N$ (see~\cite[Section~2.5]{gohberg2006invariant}).  In this case the set of Jordan subspaces has one element $\mathscr J = \mathbb C^N$. Properties of unicellular Jordan equivalence classes are demonstrated next.
\begin{theorem}
\label{thm:unicell_allpass}
Let $\mathcal G(A)$ be an element of the unicellular Jordan equivalence class $\mathbf G_A^J$. Then all graph filters $H\in \mathbf G_A^J$ are all-pass.
\end{theorem}
\begin{IEEEproof}
Since $A$ is unicellular, it has a single Jordan chain $v_1,\dots,v_N$ of length $N$. Consider a graph signal~$s$ over graph~$\mathcal G(A)$, and let $\widetilde s$ represent the coordinate vector of~$s$ in terms of the basis $\{v_i\}_{i=1}^N$. Then the spectral decomposition of signal~$s$ is given by
\begin{equation}
\label{eq:unicell}
s = \widetilde s_1 v_1 + \cdots \widetilde s_N v_N = \widehat s;
\end{equation}
that is, the unique projection of $s$ onto the spectral component $\mathscr J = \mathbb C^N$ is itself. Therefore, $\mathcal G(A)$ acts as an all-pass filter. Moreover,~\eqref{eq:unicell} holds for all graphs in Jordan equivalence class~$\mathbf G_A^J$.
\end{IEEEproof}
In addition to the all-pass property of unicellular graph filters, unicellular isomorphic graphs are also Jordan equivalent, as proven next.
\begin{theorem}
\label{thm:unicell_isomorphic}
Let $\mathcal G(A),\mathcal G(B)\in\mathbf G_A^I$ where $A$ is a unicellular matrix. Then $\mathcal G(A),\mathcal G(B)\in\mathbf G_A^J$.
\end{theorem}
\begin{IEEEproof}
Since $\mathcal G(A)$ and $\mathcal G(B)$ are isomorphic, Jordan normal forms $J_A = J_B$. Therefore, $B$ is also unicellular, so $\mathbf J_A = \mathbf J_B = \{\mathbb C^N\}$. By Definition~\ref{def:equivgraph}, $\mathcal G(A),\mathcal G(B)\in\mathbf G_A^J$.
\end{IEEEproof}
The dual basis of $V$ can also be used to construct graphs in the Jordan equivalence class of unicellular~$A$.
\begin{theorem}
\label{thm:unicell_dualbasis}
Denote by $V$ an eigenvector matrix of unicellular $A\in\mathbb C^{N\times N}$ and $W=V^{-H}$ is the dual basis. Consider decompositions $A = VJV^{-1}$ and $A_W = WJW^{-1}$. Then $\mathcal G(A_W)\in\mathbf G_A^J$.
\end{theorem}
\begin{IEEEproof}
Matrices $A$ and $A_W$ have the same Jordan normal form by definition. Since there is only one Jordan block, both matrices have a single Jordan subspace $\mathbb C^N$. By Definition~\ref{def:equivgraph}, $\mathcal G(A_W)$ and $\mathcal G(A)$ are Jordan equivalent.
\end{IEEEproof}
The next theorem characterizes the special case of graphs in the Jordan equivalence class that contains~$\mathcal G(J)$ with adjacency matrix equal to Jordan block $J=J(\lambda)$.
\begin{theorem}
\label{thm:unicell_uppertri}
Denote by $J=J(\lambda)$ is the $N\times N$ Jordan block~\eqref{eq:jordanblock} for eigenvalue $\lambda$. Then $\mathcal G(A)\in\mathbf G_J^J$  if~$A\in\mathbb C^{N\times N}$ is upper triangular  with diagonal entries $\lambda$ and nonzero entries on the first off-diagonal.
\end{theorem}
\begin{IEEEproof}
Consider upper triangular matrix $A=[a_{ij}]$ with diagonal entries~$a_{11} = \cdots=a_{NN}$ and nonzero elements on the first off-diagonal.  By~\cite[Example~10.2.1]{gohberg2006invariant},~$A$ has the same invariant subspaces as $J=J(\lambda)$, which implies~$\mathbf J_J= \mathbf J_A = \{\mathbb C^N\}$. Therefore, the Jordan normal form of~$A$ is the Jordan block~$J_A = J(a_{11})$. Restrict the diagonal entries of $A$ to $\lambda$ so $J_A = J$. Then, $\mathcal G(J),\mathcal G(A)\in\mathbf G_J^J$ by Definition~\eqref{def:equivgraph}.
\end{IEEEproof}

Figure~\ref{fig:equivgraphs_unicell} shows graph structures that are in the same unicellular Jordan equivalence class by Theorem~\ref{thm:unicell_uppertri}. 
In addition, the theorem implies that it is sufficient to determine the GFT of unicellular~$A$ by replacing~$\mathcal G(A)\in\mathbf G_J^J$ with~$\mathcal G(J)$, where~$J$ is a single $N\times N$ Jordan block. That is, without loss of generality,~$\mathcal G(A)$ can be replaced with a directed chain graph with possible self-edges and  the eigenvector matrix~$V=I$ chosen to compute the GFT of a graph signal.

%
\textbf{Remark on invariant spaces.} Example 10.2.1 of~\cite{gohberg2006invariant} shows that a matrix~$A\in\mathbb C^{N\times N}$ having upper triangular entries  with constant diagonal entries $a$ and nonzero entries on the first off-diagonal is both necessary and sufficient for $A$ to have the same invariant subspaces as $N\times N$ Jordan block $J=J(\lambda)$ (i.e.,  $\mathrm{Inv}(J) = \mathrm{Inv}(A)$, where $\mathrm{Inv}(\cdot)$ represents the set of invariant spaces of a matrix). If $a=\lambda$, Definition~\ref{def:equivgraph} can be applied, which yields $\mathcal G(A)\in\mathbf G_J^J$.

On the other hand, consider a unicellular matrix $B$ such that its eigenvector is not in the span of a canonical vector, e.g.,
\renewcommand{\arraystretch}{1.2}
\begin{equation}
\label{eq:unicellexample_not01}
\setlength\arraycolsep{12pt}
B =
\begin{bmatrix}
\frac{1}{2} & -\frac{1}{2} & \frac{1}{2} & \frac{1}{2}\\
\frac{1}{2} & -\frac{1}{2} & -\frac{1}{2} & -\frac{1}{2} \\
0 & 0 & \frac{1}{2} & -\frac{1}{2} \\
0 & 0 & \frac{1}{2} & -\frac{1}{2}
\end{bmatrix}
\end{equation}
with Jordan normal form $J(0)$. Since the span of the eigenvectors of $J(0)$ and $B$ are not identical, $\mathrm{Inv}(J(0))\neq\mathrm{Inv}(B)$. However, by Definition~\ref{def:equivgraph}, $\mathcal G(B)$ is in the same class of unicellular Jordan equivalent graphs as those of Figure~\ref{fig:equivgraphs_unicell}, i.e., $\mathcal G(B)\in\mathbf G_J^J$.  In other words, for matrices~$A$ and $B$ with the same Jordan normal forms ($J_A=J_B$), Jordan equivalence, i.e., $\mathbf J_A = \mathbf J_B$, is a more general condition than $\mathrm{Inv}(A) = \mathrm{Inv} (B)$. This illustrates that graphs having adjacency matrices with equal Jordan normal forms and the same sets of invariant spaces form a proper subset of a Jordan equivalence class, as shown above in Theorem~\ref{thm:jordanequiv_sameinvariantspaces}.

\textbf{Remark on topology.} Note that replacing each nonzero element of~\eqref{eq:unicellexample_not01}  with a unit entry results in a matrix that is not unicellular. Therefore, its corresponding graph is not in a unicellular Jordan equivalence class. This observation demonstrates that topology may not determine the Jordan equivalence class of a graph.

\subsection{Two Jordan Blocks}
\label{sec:jordanequiv:twojordanblocks}
Consider $N\times N$ matrix $A$ with Jordan normal form consisting of two Jordan subspaces $\mathscr J_1 = \mathrm{span}(v_1,\dots,v_{r_1})$ and $\mathscr J_2 = \mathrm{span}(v_{r_1+1},\dots,v_{r_2})$ of dimensions $r_1>1$ and $r_2= N-r_1$ and corresponding eigenvalues $\lambda_1$ and $\lambda_2$, respectively. The spectral decomposition of signal~$s$ over~$\mathcal G(A)$ yields
\begin{align}
\label{eq:unicell2}
s &= \underbrace{\widetilde s_1 v_1 + \cdots + \widetilde s_{r_1} v_{r_1}}_{\widehat s_1} + \underbrace{\widetilde s_{r_1 +1} v_{r_1+1} + \cdots +\widetilde s_N v_N}_{\widehat s_2} \\
&=  \widehat s_1 + \widehat s_2.
\end{align}

Spectral components $\widehat s_1$ and $\widehat s_2$ are the unique projections of $s$ onto the respective Jordan subspaces. By Example~6.5.4 in~\cite{lancaster1985},  a Jordan basis matrix~$X$ can be chosen for $A=VJV^{-1}$ such that $X = VU$, where $U$ commutes with $J$ and has a particular form as follows.

If $\lambda_1\neq \lambda_2$, then $U = \mathrm{diag}(U_1,U_2)$, where $U_i$, $i=1,2$, is an $r_i\times r_i$ upper triangular Toeplitz matrix; otherwise,~$U$ has form
\begin{equation}
\label{eq:toeplitz}
\setlength\arraycolsep{12pt}
U = \mathrm{diag}(U_1,U_2) + \begin{bmatrix}
\underline 0& U_{12}\\ U_{21}& \underline 0
\end{bmatrix}
\end{equation}
where~$U_i$ is an $r_i\times r_i$ upper triangular Toeplitz matrix and~$U_{12}$ and $U_{21}$ are extended upper triangular Toeplitz matrices as in Theorem~12.4.1 in~\cite{lancaster1985}. Thus, all Jordan bases of $A$ can be obtained by transforming eigenvector matrix $V$ as $X = VU$.

A corresponding theorem to Theorem~\ref{thm:unicell_uppertri} is presented to characterize Jordan equivalent classes when the Jordan form consists of two Jordan blocks. The reader is directed to Sections~10.2 and~10.3 in~\cite{gohberg2006invariant} for more details. The following definitions are needed. Denote $p\times p$ upper triangular Toeplitz matrices
$T_{r_2}\left(b_1,\dots,b_{r_2}\right)$ of form
\begin{align}
\label{eq:gohberg_T}
\setlength\arraycolsep{12pt}
 &T_{p}\left(b_1,\dots,b_{p}\right) =
 \begin{bmatrix}
 b_1 & b_2 &\cdots & b_{p -1} & b_{p}\\
 0 & b_1 & \ddots  &b_{p-2} & b_{p-1}\\
 \vdots &  \vdots & \ddots &\ddots   &\vdots \\
  0 & 0 & \cdots  & b_1 & b_2\\
 0 & 0 &  \cdots & 0 & b_1\\
 \end{bmatrix},
\end{align}
and define $q\times q$ upper triangular matrix for some $q>p$
 \renewcommand{\arraystretch}{.7}
 \setlength\arraycolsep{3pt}
\begin{align}
\label{eq:gohberg_R}
 &R_q\left(b_1,\dots,b_{p}; F\right) = \nonumber\\
 &\begin{bmatrix}
 b_1  &\cdots &b_{p} & f_{11} & f_{12}  &\cdots  f_{1,q-p -1} & f_{1,q-p}\\
 0 & b_1 & \cdots &b_{p} & f_{22}  &\cdots  f_{2,q-p -1} & f_{2,q-p}\\
 \vdots &  \vdots & \ddots    &&\ddots    &\vdots & \vdots\\
  0 & 0  &  & &\cdots & b_{p}  &  f_{q-p,q-p}\\
  0 & 0  &  & & \cdots& b_{p-1}  &  b_{p}\\
 \vdots &  \vdots   &  & &\vdots & \vdots  & \vdots \\
  0 & 0   & \cdots & &  0 & b_1 & b_2\\
 0 & 0   & \cdots & &   0& 0 & b_1\\
 \end{bmatrix}
\end{align}
where $F = [f_{ij}]$ is a $(q-p)\times(q-p)$ upper triangular matrix and $b_i\in\mathbb C$, $i=1,\dots,p$. The theorems are presented below.
\begin{theorem}
\label{thm:twoblocks_uppertri_difflambda}
Consider $A = \mathrm{diag}(A_1,A_2)$ where each matrix $A_i$, $i=1,2$, is upper triangular with diagonal elements $\lambda_i$ and nonzero elements on the first off-diagonal. Let $\lambda_1 \neq \lambda_2$. Then $\mathcal G(A)$ is Jordan equivalent to the graph with adjacency matrix $J = \mathrm{diag}(J_{r_1}(\lambda_1),J_{r_2}(\lambda_2))$ where $J_{r_i}(\lambda_i)$ is the $r_i\times r_i$ Jordan block for eigenvalue $\lambda_i$.
\end{theorem}
\begin{IEEEproof}
By Theorem~\ref{thm:unicell_uppertri}, $\mathcal G(A_i)$ and $\mathcal G(J_{r_i}(\lambda_i))$ are Jordan equivalent for $i=1,2$ and $A_i$ upper triangular with nonzero elements on the first off-diagonal. Therefore, the Jordan normal forms of $J$ and $A$ are the same. Moreover, the set of irreducible subspaces of $J$ is the union of the irreducible subspaces  of $[J_1\  \underline 0]^T$ and $[\underline 0\  J_2]^T$, which are the same as the irreducible subspaces of $[A_1 \ \underline 0
]^T$ and $[\underline 0 \ A_2]^T$, respectively. Therefore, $\mathbf J_A = \mathbf J_J$, so $\mathcal G(A)$ and $\mathcal G(J)$ are Jordan equivalent.
\end{IEEEproof}
\begin{theorem}
\label{thm:twoblocks_uppertri_samelambda}
Consider $A = \mathrm{diag}(A_1,A_2)$ where $A_1 = U_{r_1}(\lambda,b_1,\dots,b_{r_2},F)$ and $A_2 = T_{r_2}(\lambda,b_1,\dots,b_{r_2})$, $r_1\geq r_2$.  Then $\mathcal G(A)$ is Jordan equivalent to the graph with adjacency matrix $J = \mathrm{diag}(J_{r_1}(\lambda),J_{r_2}(\lambda))$ where $J_{r_i}(\lambda)$ is the $r_i\times r_i$ Jordan block for eigenvalue $\lambda$.
\end{theorem}
\begin{IEEEproof}
By Lemma~10.3.3 in~\cite{gohberg2006invariant}, $A$ with structure as described in the theorem have the same invariant subspaces as $J = \mathrm{diag}(J_{r_1}(\lambda),J_{r_2}(\lambda))$. Therefore, $A$ and $J$ have the same Jordan normal form and Jordan subspaces and so are Jordan equivalent.
\end{IEEEproof}

Theorems~\ref{thm:twoblocks_uppertri_difflambda} and~\ref{thm:twoblocks_uppertri_samelambda} demonstrate two types of Jordan equivalences that arise from block diagonal matrices with submatrices of form~\eqref{eq:gohberg_T} and~\eqref{eq:gohberg_R}. These theorems imply that computing the GFT~\eqref{eq:gft} over the block diagonal matrices can be simplified to computing the transform over the adjacency matrix of a union of directed chain graphs. That is, the canonical basis can be chosen for $V$ without loss of generality.

As for the case of unicellular transformations, it is possible to pick bases of $\mathscr J_1$ and $\mathscr J_2$ that do not form a Jordan basis of~$A$. Any two such choices of bases are related by Theorem~\ref{thm:jordanequiv_transformation}. Concretely, if $V$ is the eigenvector matrix of $A$ and $X$ is the matrix corresponding to another choice of basis, then Theorem~\ref{thm:jordanequiv_transformation} states that 
a transformation matrix~$Y$ can be found such that $X = VY$, where $Y$ is partitioned as $Y = \mathrm{diag}(Y_1, Y_2)$ with full-rank submatrices $Y_i\in\mathbb C^{r_i\times r_i}$, $i=1,2$. 
%
%
\subsection{Multiple Jordan Blocks}
\label{sec:jordanequiv:multipleblocks}
This section briefly describes a special case of Jordan equivalence classes whose graphs have adjacency matrices~$A\in\mathbb C^{N\times N}$ with~$p$ Jordan blocks, $1<p<N$.

Consider matrix $A$ with Jordan normal form~$J$ comprised of $p$ Jordan blocks and eigenvalues $\lambda_1,\dots,\lambda_k$. 
By Theorem~10.2.1 in~\cite{gohberg2006invariant}, there exists an upper triangular~$A$ with Jordan decomposition $A=VJV^{-1}$ such that $\mathcal G(A)\in\mathbf G_J^J$. Note that the elements in the Jordan equivalence class~$\mathbf G_J^J$ of $\mathcal G(J)$ are useful since signals over a graph in this class can be computed with respect to the canonical basis with eigenvector matrix $V=I$.
Theorem~\ref{thm:multicell_V} characterizes the possible eigenvector matrices~$V$ such that $A = VJV^{-1}$ allows $\mathcal G(A)\in\mathbf G_J^J$.
\begin{theorem}
\label{thm:multicell_V}
Let $A = VJV^{-1}$ be the Jordan decomposition of $A\in\mathbb C^{N\times N}$ and $\mathcal G(A)\in\mathbf G_J^J$. Then $V$ must be an invertible block diagonal matrix.
\end{theorem}
\begin{IEEEproof}
Consider $\mathcal G(J)$ with eigenvector matrix $I$. By Theorem~\ref{thm:jordanequiv_transformation},~$\mathcal G(A)\in\mathbf G_J^J$ implies
\begin{equation}
V = IY =Y
\end{equation}
where $Y$ is an invertible block diagonal matrix.
\end{IEEEproof}

The structure of $V$ given in Theorem~\ref{thm:multicell_V} allows a characterization of graphs in the Jordan equivalence class~$\mathbf G_J^J$ with the dual basis of $V$ as proved in Theorem~\ref{thm:multicell_dualbasis}.
\begin{theorem}
\label{thm:multicell_dualbasis}
Let $\mathcal G(A)\in\mathbf G_J^J$, where $A$ has Jordan decomposition $A = VJV^{-1}$  and $W=V^{-H}$ is the dual basis of $V$. If $A_W = WJW^{-1}$, then $\mathcal G(A_W)\in\mathbf G_J^J$.
\end{theorem}
\begin{IEEEproof}
By Theorem~\ref{thm:multicell_V}, $V$ is block diagonal with invertible submatrices $V_i$. Thus, $W = V^{-H}$ is block diagonal with submatrices $W_{i} = V_{i}^{-H}$. By Theorem~\ref{thm:jordanequiv_transformation}, $W$ is an appropriate eigenvector matrix such that, for $A_W = W JW^{-1}$, $\mathcal G(A_W)\in\mathbf G_J^J$.
\end{IEEEproof}

\textbf{Relation to graph topology.} Certain types of matrices have Jordan forms that can be deduced from their graph structure. For example,~\cite{cardon2011jordan} and~\cite{nina2013jordan} relate the Jordan blocks of certain adjacency matrices to a decomposition of their graph structures into unions of cycles and chains. Applications where such graphs are in use would allow a practitioner to determine the Jordan equivalence classes (assuming the eigenvalues can be computed) and  potentially choose a different matrix in the class for which the GFT  can be computed more easily. Sections~\ref{sec:jordanequiv:onejordanblock} and~\ref{sec:jordanequiv:multipleblocks} show that working with unicellular matrices and matrices in Jordan normal form  permits the choice of the canonical basis. In this way, for matrices with  Jordan blocks of size greater than one, finding a spanning set for each Jordan subspace may be more efficient than attempting to compute the Jordan chains.  Nevertheless, relying on graph topology is not always possible. Such an example was presented in Section~\ref{sec:jordanequiv:onejordanblock} with adjacency matrix~\eqref{eq:unicellexample_not01}.

\textbf{Relation to algebraic signal processing.} The emergence of Jordan equivalence from the graph Fourier transform~\eqref{eq:gft} is related to algebraic signal processing and the signal model $(\mathcal A, \mathcal M, \Phi)$, where $\mathcal A$ is a signal algebra corresponding to the filter space,~$\mathcal M$ is a module of~$\mathcal A$ corresponding to the signal space, and $\Phi: V \rightarrow \mathcal M$ is a bijective linear mapping that generalizes the~$z$-transform~\cite{puschel2008algebraic_foundation,puschel2008algebraic_1dspace}. We emphasize that the GFT~\eqref{eq:gft} is tied to a basis. This is most readily seen by considering diagonal adjacency matrix $A=\lambda I$, where any basis that spans $\mathbb C^N$ defines the eigenvectors (the Jordan subspaces and spectral components) of a graph signal; that is, a matrix, even a diagonalizable matrix, may not have distinct spectral components. Similarly, the signal model $(\mathcal A, \mathcal M, \Phi)$ requires a choice of basis for module (signal space)~$\mathcal M$ in order to define the frequency response (irreducible representation) of a signal~\cite{puschel2008algebraic_foundation}. 
On the other hand, this section demonstrated the equivalence of the GFT~\eqref{eq:gft} over graphs in Jordan equivalence classes, which implies an equivalence of certain bases. This observation suggests the concept of \emph{equivalent signal models} in the algebraic signal processing framework. Just as working with graphs that are Jordan equivalent to those with adjacency matrices in Jordan normal form simplifies GFT computation, we expect similar classes of equivalent signal models for which the canonical basis can be chosen without loss of generality.


Jordan equivalence classes show that the GFT~\eqref{eq:gft} permits degrees of freedom in graph topologies. This has ramifications for total variation-based orderings of the spectral components, as discussed in the next section.
\section{Frequency Ordering of Spectral Components}
\label{sec:jordanequiv:totalvar}
This section defines a mapping of spectral components to the real line to achieve an ordering of the spectral components. This ordering can be used to distinguish generalized low and high frequencies as in~\cite{sandryhaila2014discrete}. An upper bound for a total-variation based mapping of a spectral component (Jordan subspace) is derived and generalized to Jordan equivalence classes.

The \emph{graph total variation}  of a graph signal $s\in\mathbb C^N$ is defined as~\cite{sandryhaila2014discrete}
\begin{equation}
\label{eq:TV_signal_graph}
\mathrm{TV}_G\left(s\right) =  \left\|s - A s\right\|_1.
\end{equation}
Matrix $A$ can be replaced by $A^{\mathrm{norm}} = \frac{1}{\left|\lambda_{\mathrm{max}}\right|} A$ when the maximum eigenvalue satisfies $|\lambda_{\mathrm{max}}|>0$.

Equation~\eqref{eq:TV_signal_graph} can be generalized to define the {total variation} of the Jordan subspaces of the graph shift~$A$ as described in~\cite{deriGFT2016}. 
Choose a Jordan basis of~$A$ so that~$V$ is the eigenvector matrix of~$A$, i.e., $A= VJV^{-1}$, where $J$ is the Jordan form of~$A$. Partition $V$ into $N\times r_{ij}$ submatrices $V_{ij}$ whose columns are a Jordan chain of (and thus span) the $j$th Jordan subspace~$\mathscr J_{ij}$ of eigenvalue~$\lambda_i$, $i=1,\dots, k\leq N$, $j = 1,\dots,g_{i}$.
Then the (graph) total variation of $V_{ij}$ is defined as~\cite{deriGFT2016}
\begin{equation}
\label{eq:TV_singlecomp}
\mathrm{TV}_G\left(V_{ij}\right) =  \left\|V_{ij} - A V_{ij} \right\|_1,
\end{equation}
where $\|\cdot\|_1$ represents the induced L1 matrix norm (equal to the maximum absolute column sum). 

Theorem~\ref{thm:TVisomorphic} shows that the graph total variation of a spectral component is invariant to a relabeling of the graph nodes; that is, the total variations of the spectral components for graphs in the same isomorphic equivalence class as defined in Section~\ref{sec:gft:isomorphic} are equal.
\begin{theorem}
\label{thm:TVisomorphic}
Let $A,B\in\mathbb C^{N\times N}$ and $\mathcal G(B)\in \mathbf G^I_A$, i.e., $\mathcal G(B)$ is isomorphic to $\mathcal G(A)$. Let $V_{A,ij}\in\mathbb C^{N\times_{r_{ij}}}$ be a Jordan chain of matrix $A$ and $V_{B,ij}\in\mathbb C^{N\times_{r_{ij}}}$ the corresponding Jordan chain of $B$. Then \begin{equation}\mathrm{TV}_G(V_{A,ij}) = \mathrm{TV}_G(V_{B,ij}).\end{equation}
\end{theorem}
\begin{IEEEproof}
Since $\mathcal G(A)$ and $\mathcal G(B)$ are isomorphic, there exists a permutation matrix $T$ such that $B =  TAT^{-1}$ and the eigenvector matrices $V_A$ and $V_B$ of $A$ and $B$, respectively, are related by $V_B = TV_A$. Thus, the Jordan chains are related by $V_{B,ij} = TV_{A,ij}$. By~\eqref{eq:TV_singlecomp},
\begin{align}
\mathrm{TV}\left(V_B\right) & = \left\| V_{B,ij} - B V_{B,ij} \right\|_1\\
&= \left\| TV_{A,ij} - \left(TAT^{-1}\right) TV_{A,ij} \right\|_1\\
&= \left\| TV_{A,ij} - TAV_{A,ij} \right\|_1\\
&= \left\| T\left(V_{A,ij} - AV_{A,ij}\right) \right\|_1\\
&= \left\| V_{A,ij} - AV_{A,ij}\right\|_1 \label{eq:TVstep_isomorphic_colsum}\\
&= \mathrm{TV}\left(V_A\right),
\end{align}
where~\eqref{eq:TVstep_isomorphic_colsum} holds because the maximum absolute column sum of a matrix is invariant to a permutation on its rows.
\end{IEEEproof}

Theorem~\ref{thm:TVisomorphic} shows that the graph total variation is invariant to a node relabeling, which implies that an ordering of the total variations of the frequency components is also invariant.

Reference~\cite{deriGFT2016} demonstrates that each eigenvector submatrix corresponding to a Jordan chain can be normalized. This is stated as a property below:
\begin{property}
\label{prop:norm1Jordanchain}
The eigenvector matrix~$V$ of adjacency matrix~$A\in\mathbb{C}^{N\times N}$ can be chosen so that each Jordan chain represented by the eigenvector submatrix $V_{ij}\in\mathbb C^{N\times r_{ij}}$ satisfies $\left\|V_{ij}\right\|_1 = 1$; i.e., $\left\|V \right\|_1 = 1$ without loss of generality.
\end{property}
It is assumed that the eigenvector matrices are normalized as in Property~\ref{prop:norm1Jordanchain} for the remainder of the section.

Furthermore,~\cite{deriGFT2016} shows that~\eqref{eq:TV_singlecomp} can be written as
\begin{align}
&\mathrm{TV}_G\left(V_{ij}\right) = \left\|V_{ij}\left(I_{r_{ij}} - J_{ij} \right) \right\|_1\label{eq:TV_beforeineq}\\
&= \max_{i=2,\dots,r_{ij}} \left\{\left|1-\lambda\right| \left\|v_1\right\|_1, \left\|\left(1-\lambda\right) v_i - v_{i-1}\right\|_1\right\}.\label{eq:TV_jordanchainvecs}
\end{align}
and establishes the upper bound for the total variation of spectral components as
\begin{equation}\label{eq:TV_ineqeval} \mathrm{TV}_G(V_{ij})\leq \left|1-\lambda_i \right|+1.\end{equation}

Equations~\eqref{eq:TV_beforeineq},~\eqref{eq:TV_jordanchainvecs}, and~\eqref{eq:TV_ineqeval} characterize the (graph) total variation of a Jordan chain by quantifying the change in a set of vectors that spans the Jordan subspace~$\mathscr J_{ij}$ when they are transformed by the graph shift~$A$. These equations, however, are dependent on a particular choice of Jordan basis.  As seen in Sections~\ref{sec:jordanequiv:onejordanblock},~\ref{sec:jordanequiv:twojordanblocks}, and~\ref{sec:jordanequiv:multipleblocks}, defective graph shift matrices belong to Jordan equivalence classes that contain more than one element, and the GFT of a signal is the same over any graph in a given Jordan equivalence class. Furthermore, for any two graphs $\mathcal G(A),\mathcal G(B)\in\mathbf G_A^J$,~$A$ and~$B$ have Jordan bases for the same Jordan subspaces, but the respective total variations of the spanning Jordan chains as computed by~\eqref{eq:TV_beforeineq} may be different. Since it is desirable to be able  to order spectral components in a manner that is invariant to the choice of Jordan basis, we derive here a definition of the total variation of a spectral component of $A$ in relation to the Jordan equivalence class $\mathbf G_A^J$.

\textbf{Class total variation.} Let $\mathcal G(B)$ be an element in Jordan equivalence class~$\mathbf G_A^J$ where $B$ has Jordan decomposition $B = VJ V^{-1}$.  Let the columns of eigenvector submatrix $V_{ij}$ span the Jordan subspace~$\mathscr J_{ij}$ of $A$. Then the \emph{class total variation} of spectral component $\mathscr J_{ij}$  is defined as the supremum of the graph total variation of $V_{ij}$ over the Jordan equivalence class (for all $\mathcal G(B) \in\mathbf G_A^J$):
\begin{align}
\label{eq:TV_J}
\mathrm{TV}_{\mathbf G_A^J}\left(\mathscr J_{ij}\right) & = \sup_{\substack{\mathcal G\left(B\right)\in\mathbf G_A^J\\ B = VJV^{-1}\\ \mathrm{span}\left\{V_{ij}\right\} = \mathscr J_{ij}\\ \left\| V_{ij} \right\|_1 =1 }} \mathrm{TV}_G\left( V_{ij}\right).
\end{align}

\begin{theorem}
\label{thm:TVclassisomorphism}
Let $A,B\in\mathbb C^{N\times N}$ and $\mathcal G(B)\in\mathbf G^I_A$. Let $V_A$ and $V_B$ be the respective eigenvector matrices with Jordan subspaces $\mathscr J_{A,ij}  = \mathrm{span}\{V_{A,ij}\}$ and $\mathscr J_{B,ij}  = \mathrm{span}\{V_{B,ij}\}$ spanned by the $j$th Jordan chain of eigenvalue~$\lambda_i$. Then $\mathrm{TV}_{\mathbf G^J_A}(\mathscr J_{A,ij}) = \mathrm{TV}_{\mathbf G^J_B}(\mathscr J_{B,ij})$.
\end{theorem}
\begin{IEEEproof}
Let $V ^\ast_A$ denote the eigenvector matrix corresponding to $\mathcal G(A^\ast)\in\mathbf G_A^J$ that maximizes the class total variation of Jordan subspace $\mathscr J_{A,ij}$; i.e.,
\begin{equation}
\mathrm{TV}_{\mathbf G^J_A}\left(\mathscr J_{A,ij}\right) =  \mathrm{TV}_{G}\left(V^\ast_{A,ij}\right).
\end{equation}
Similarly, let $V ^\ast_B$ denote the eigenvector matrix corresponding to $\mathcal G(B^\ast)\in\mathbf G_B^J$ that maximizes the class total variation of Jordan subspace $\mathscr J_{B,ij}$, or
\begin{equation}
\mathrm{TV}_{\mathbf G^J_B}\left(\mathscr J_{B,ij}\right) =  \mathrm{TV}_{G}\left(V^\ast_{B,ij}\right).
\end{equation}

Since $\mathcal G(A)$ and $\mathcal G(B)$ are isomorphic, Theorem~\ref{thm:jordanequiv_isomorphism} implies that there exists $\mathcal G(B')\in\mathbf G_B^J$ such that $B' = TA^\ast T^{-1}$; i.e., $V_{B'} = TV^\ast_{A}$ where $V_{B'}$ is an eigenvector matrix of $B'$. By the class total variation definition~\eqref{eq:TV_J}, $\mathrm{TV}_G(V_{B',ij})\leq \mathrm{TV}_G(V^\ast_{B,ij})$. Applying Theorem~\ref{thm:TVisomorphic} to isomorphic graphs $\mathcal G(A^\ast)$ and $\mathcal G(B')$ yields
\begin{equation}
\label{eq:classisomorph1}
\mathrm{TV}_{G}\left(V^\ast_{A,ij}\right) = \mathrm{TV}_{G}\left(V_{B',ij}\right)\leq \mathrm{TV}_G\left(V^\ast_{B,ij}\right).
\end{equation}

Similarly, by Theorem~\ref{thm:jordanequiv_isomorphism}, there exists $\mathcal G(A')\in\mathbf G_A^J$ such that $B^\ast = TA'T^{-1}$, or $V_{B}^\ast = TV_{A'}$ where $V_{A'}$ is an eigenvector matrix of $A'$. Apply~\eqref{eq:TV_J} and Theorem~\ref{thm:TVisomorphic} again to obtain
\begin{equation}
\label{eq:classisomorph2}
\mathrm{TV}_{G}\left(V^\ast_{A,ij}\right)\geq \mathrm{TV}_{G}\left(V_{A',ij}\right) = \mathrm{TV}_{G}\left(V^\ast_{B,ij}\right).
\end{equation}
Equations~\eqref{eq:classisomorph1} and~\eqref{eq:classisomorph2} imply that $\mathrm{TV}_{G}\left(V^\ast_{A,ij}\right) = \mathrm{TV}_{G}\left(V^\ast_{B,ij}\right)$, or
\begin{equation}
\mathrm{TV}_{\mathbf G^J_A}\left(\mathscr J_{A,ij}\right)=\mathrm{TV}_{\mathbf G^J_B}\left(\mathscr J_{B,ij}\right).
\end{equation}
\end{IEEEproof}

Theorem~\ref{thm:TVclassisomorphism} shows that the class total variation of a spectral component is invariant to a relabeling of the nodes. This is significant because it means that an ordering of the spectral components by their class total variations is invariant to node labels.

Next, the significance of the class total variation~\eqref{eq:TV_J} is illustrated for adjacency matrices with diagonal Jordan form, one Jordan block, and multiple Jordan blocks.

\textbf{Diagonal Jordan Form.} Section~\ref{sec:jordanequiv:diagonaljordanform} shows that a graph shift~$A$ with diagonal Jordan form is the single element of its Jordan equivalence class $\mathbf G_A^J$. This yields the following result.
\begin{theorem}
\label{thm:TV_J_diagonal}
Let $\mathcal G(A)$ have diagonalizable adjacency matrix $A$ with eigenvectors $v_1,\dots,v_N$. Then the class total variation of the spectral component $\mathscr J_i$, $i=1,\dots, N$, of $A$ satisfies (for $\left\|v_i\right\|=1$)
\begin{equation}
\label{eq:TV_J_diagonal}
\mathrm{TV}_{\mathbf G_A^J}\left(\mathscr J_i\right) = \left|1- \lambda_i\right|.
\end{equation}
\end{theorem}
\begin{IEEEproof}
Each spectral component~$\mathscr J_i$ of~$A$ is the span of eigenvector $v_i$ corresponding to eigenvalue~$\lambda_i$. The class total variation of $\mathscr J_i$ is then
\begin{align}
\mathrm{TV}_{\mathbf G_A^J}\left(\mathscr J_{i}\right) & = \sup_{\substack{\mathcal G\left(B\right)\in\mathbf G_A^J\\ B = VJV^{-1}\\ \mathrm{span}\left\{v_{i}\right\}= \mathscr J_{i}\\ \left\| v_{i} \right\|_1 =1 }} \mathrm{TV}_G\left( v_{i}\right)\\
& = \mathrm{TV}_G\left(v_i\right)\\
&= \left\|v_i - Bv_i\right\|_1 \hspace{1cm} \text{(by \eqref{eq:TV_singlecomp})}\\
&= \left\|v_i - \lambda_i v_i\right\|_1\\
&= \left|1 - \lambda_i\right| \left\|v_i \right\|_1\\
&= \left|1 - \lambda_i\right|.
\end{align}
\end{IEEEproof}

Theorem~\ref{thm:TV_J_diagonal} is consistent with the total variation result for diagonalizable graph shifts in~\cite{sandryhaila2014discrete}. Next, the class total variation for defective graph shifts is characterized.

\textbf{One Jordan block.} Consider the graph shift~$A$ with a single spectral component $\mathscr J = \mathbb C^N$ and Jordan form $J=J(\lambda)$. The next theorem proves that the total variation of $\mathscr J$ attains the upper bound~\eqref{eq:TV_ineqeval}.
\begin{theorem}
\label{thm:TV_J_unicell}
Consider unicellular $A\in\mathbb C^{N\times N}$ with Jordan normal form $J=J(\lambda)$. Then the class total variation of $\mathbf G_A^J$ is $\left|1-\lambda\right|+1$.
\end{theorem}
\begin{IEEEproof}
Graph $\mathcal G(A)$ is Jordan equivalent to $\mathcal G(J)$ since $A$ is unicellular. Therefore, the GFT of a graph signal can be computed over $\mathcal G(J)$ by choosing the the canonical  vectors ($V=I$) as the Jordan basis, as shown in~\eqref{eq:unicell}.  By~\eqref{eq:TV_jordanchainvecs}, the maximum of $\left|1 - \lambda \right| \left\|  v_1\right\|_1$ and $\left\| \left|1 - \lambda \right|   v_i - v_{i-1}\right\|_1$ for $i=2,\dots,N$ needs to be computed. The former term equals $\left|1-\lambda\right|$ since $v_1$ is the first canonical vector. The latter term has form
\begin{align}
\left\|\,\, \left|1 - \lambda \right|   v_i - v_{i-1}\right\|_1 & = \left\| \begin{bmatrix}
\underline 0 \\-1\\ \left|1 - \lambda \right|\\ \underline 0
\end{bmatrix}\right\|_1\\
&= 1 + \left| 1 - \lambda \right|,
\end{align}
Since  $\left|1-\lambda\right|  + 1 >\left|1-\lambda\right|$, $\mathrm{TV}_G(I) =1 + \left| 1 - \lambda \right|$. Therefore,~\eqref{eq:TV_ineqeval} holds with equality, so the class total variation of $\mathscr J = \mathbb C^N$ satisfies
\begin{align}
\label{eq:TV_J_unicell}
\mathrm{TV}_{\mathbf G_A^J}\left(\mathscr J_{i}\right) & = \sup_{\substack{\mathcal G\left(B\right)\in\mathbf G_A^J\\ B = VJV^{-1}\\ \mathrm{span}\left\{V\right\}= \mathscr J =\mathbb C^N\\ \left\| V \right\|_1 =1 }} \mathrm{TV}_G\left( V\right)\\
& = \mathrm{TV}_G\left(I\right)\\
&= \left|1 - \lambda\right| +1.
\end{align}

\end{IEEEproof}

\textbf{Multiple Jordan blocks.} Theorem~\ref{thm:TV_J_multicell} proves that graphs in the Jordan equivalence class $\mathbf G_J^J$ where $J$ is in Jordan normal form attains the bound~\eqref{eq:TV_ineqeval}.
\begin{theorem}
\label{thm:TV_J_multicell}
Let $\mathcal G(A)\in\mathbf G_J^J$ where $J$ is the Jordan normal form of $A$ and $\mathbf J_A = \{\mathscr J_{ij} \}_{ij}$ for $i=1,\dots,k$, $j=1,\dots,g_i$. Then the class total variation of $\mathscr J_{ij}$ is $\left|1-\lambda_i \right| +1$.
\end{theorem}
\begin{IEEEproof}
Since $\mathcal G(A)\in\mathbf G_J^J$, the GFT can be computed over $\mathcal G(J)$ with eigenvector matrix $V=I$. Then each $V_{ij} = I_{r_{ij}}$ that spans $\mathscr J_{ij}$ has total variation
\begin{align}
\mathrm{TV}_G\left(I_{ij}\right) &= \left\| I_{r_{ij}} - J_{ij} \right\|_1\\
&=\left|1 -\lambda_i\right| +1 \hspace{1cm}\text{(by~\eqref{eq:TV_ineqeval})}.
\end{align}
Therefore,
\begin{align}
\label{eq:TV_J_multicell}
\mathrm{TV}_{\mathbf G_J^J}\left(\mathscr J_{i}\right) & = \sup_{\substack{\mathcal G\left(B\right)\in\mathbf G_A^J\\ B = VJV^{-1}\\ \mathrm{span}\left\{V_{ij}\right\}= \mathscr J_{ij} \\ \left\| V_{ij} \right\|_1 =1 }} \mathrm{TV}_G\left( V_{ij}\right)\\
& = \mathrm{TV}_G\left(I_{r_{ij}}\right)\\
&= \left|1 - \lambda_i\right| +1.
\end{align}
\end{IEEEproof}

Although the total variation upper bound may not be attained for a general graph shift~$A$, choosing this bound as the ordering function provides  a useful standard for comparing  spectral components for all graphs in a Jordan equivalence class. The ordering proceeds as follows:
\begin{enumerate}
\item Order the eigenvalues $\lambda_1,\dots,\lambda_k$ of $A$ by increasing $\left|1-\lambda_i\right| +1$ (from low to high total variation).
\item Permute submatrices $V_{ij}$ of eigenvector matrix $V$ to respect the total variation ordering.
\end{enumerate}
Since the ordering is based on the class total variation~\eqref{eq:TV_J}, it is invariant to the particular choice of Jordan basis for each nontrivial Jordan subspace. Such an ordering can be used to study low frequency and high frequency behaviors of graph signals; see also~\cite{sandryhaila2014discrete}.
\section{Example}
\label{sec:example}
This section illustrates the Jordan equivalence classes of Section~\ref{sec:jordanequiv} and total variation ordering of Section~\ref{sec:jordanequiv:totalvar} on the $10\times 10$ matrix example
\arraycolsep=3pt
\renewcommand{\arraystretch}{.6}
\begin{equation}
A=\begin{bmatrix}
0 & 0 & 0 &-2 & 0 & -3& 0 & 0 & 0 & 0\\
0 & 0& 0& 0 & 0& 0& 0& 1& 0& 0\\
5 & 0& 0& 0 & 0& 0& 2& 0& 0& 0\\
0 & 0& 0& 0 & 6& 0& 0& 0& 0& 0\\
0 & 0& 0& 0 & 0& 0& 0& 1& 0& 0\\
0 & 0& 0& 0 & 0& 0& 0& 0& -2& 0\\
0 & 0& 0& 0 & 0& 0& 0& 0& 0& 3\\
0 & 0& 0& 0 & 0& 0& 0& 4& 0& 0\\
0 & 1& 0& 0 & 0& 0& 0& 0& 0& 0\\
0 & 0& -1& 0 & 0& 0& 0& 0& 0& 0
\end{bmatrix}.
\end{equation}
The Jordan normal form of $A$ is
\begin{equation}
\label{eq:exampleJ}
J=\mathrm{diag}\left(4,\sqrt[3]{-6}\omega,\sqrt[3]{-6}\omega^2,\sqrt[3]{-6}, J_4(0),J_2(0)\right),
\end{equation}
where $\omega  = \exp(2\pi j/3)$ and $J_4(0)$ and $J_2(0)$ are $4\times 4$ and $2\times 2$ Jordan blocks corresponding to eigenvalue zero, respectively.

\begin{figure}[tb]
\hspace{-.5cm}
\includegraphics[width=1.1\linewidth]{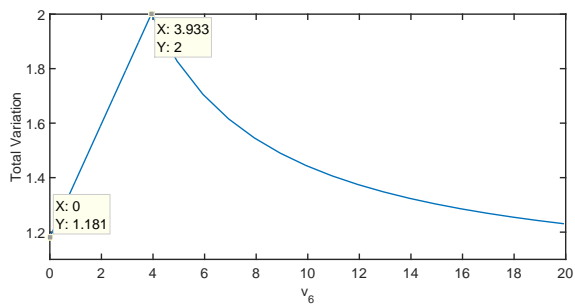}
\caption{\small Total variation of the spectral component of $J_2(0)$ for the example in Section~\ref{sec:example} with respect to generalized eigenvector component~$v_6$. The data points (gray squares) show total variation $1.181$ when $v_6=0$ and $2$ when $v_6=\frac{59}{15}$.}
\label{fig:totalvariation}
\vspace{-.10cm}
\end{figure}

\textbf{Total variation.} Possible Jordan chains for the Jordan block $J_2(0)$ and their respective total variations~\eqref{eq:TV_singlecomp} are computed. By applying the recurrence equation~\eqref{eq:jordanchain} with $\lambda=0$, the following eigenvector submatrices with columns that span potential Jordan subspaces corresponding to $J_2(0)$ in~\eqref{eq:exampleJ} are obtained:
\arraycolsep=3pt
\renewcommand{\arraystretch}{1.2}
\begin{align}
V_1&=\begin{bmatrix}
 -2			&0			&0&3			&0 			&-2				&5			&0&0			&0\\
 0			&0			&0&1 		   &\frac{1}{2}	&0 	&0			&0&1			&\frac{5}{3}
\end{bmatrix}^T,\label{eq:ex:V1}\\
V_2&=\begin{bmatrix}
 -2			&0			&0&3			&0 			&-2				&5			&0&0			&0\\
 0			&0			&0&-\frac{1}{2} &\frac{1}{2}&1 				&0			&0&1			&\frac{5}{3}
\end{bmatrix}^T,\label{eq:ex:V2}\\
V_3&=\begin{bmatrix}
-2			&0			&0&3			&0 			&-2				&5			&0&0			&0\\
0			&0			&0&-\frac{49}{10} &\frac{1}{2}&\frac{59}{15} 				&0			&0&1			&\frac{5}{3}
\end{bmatrix}^T.\label{eq:ex:V3}
\end{align}

Normalizing these matrices by their L1 norm as specified by Property~\ref{prop:norm1Jordanchain} in Section~\ref{sec:jordanequiv:totalvar}, the resulting total variations~\eqref{eq:TV_singlecomp} are
\begin{align}
\mathrm{TV}_{G}(V_1) & = 1.181\\
\mathrm{TV}_G(V_2) & = 1.389\\
\mathrm{TV}_G(V_3) & = 2.
\end{align}
These results show that the degrees of freedom in the Jordan chain recurrence~\eqref{eq:jordanchain} can lead to fluctuating total variations of the spectral components.

We compare these results to the upper bound~\eqref{eq:TV_ineqeval}, which is $\left|\lambda -1 \right| +1 = 2$ for $\lambda =0$. Our results show that this upper bound is achieved with $V_3$~\eqref{eq:ex:V3}. In this way, the class total variation~\eqref{eq:TV_J} of the Jordan subspace~$\mathscr J_2(0)=\mathrm{span}\{V_3\}$ corresponding to  Jordan block~$J_2(0)$ is
\begin{equation}
\label{eq:ex:classtotalvar}
\mathrm{TV}_{\mathbf G_A^J}\left(\mathscr J_2(0)\right)= 2.
\end{equation}
This example shows that using the class total variation or the upper bound~\eqref{eq:TV_ineqeval} as a method of ranking the spectral components by~\eqref{eq:TV_singlecomp} removes the dependency on the choice of generalized eigenvector.

We modify $V_3$~\eqref{eq:ex:V3} by varying the sixth component~$v_6$ (and fourth component $v_4$ as $v_4 = 1 - 1.5v_6$) of the generalized eigenvector in the second column. It can be verified by~\eqref{eq:jordanchain} that such vectors are valid generalized eigenvectors. The results are shown in Figure~\ref{fig:totalvariation} with the total variation plotted versus the value of~$v_6$. The data point at $v_6=0$ corresponds to the total variation of~$V_1$~\eqref{eq:ex:V1}. The figure illustrates that the total variation has a global maximum at $v_6 = \frac{59}{15}$.

\textbf{Jordan equivalence.} It can be shown that the images of the projection matrices~\eqref{eq:spectralprojectormatrix} corresponding to $V_1$\eqref{eq:ex:V1}, $V_2$~\eqref{eq:ex:V2}, and $V_3$~\eqref{eq:ex:V3} are nonidentical; that is, each choice of Jordan basis corresponds to a different Jordan equivalence class.

Consider an alternate basis for $\mathscr J_2(0)=\mathrm{span}\{V_3\}$ provided by the columns of matrix
\arraycolsep=3pt
\renewcommand{\arraystretch}{.6}
\begin{equation}
\label{eq:widetildeV1}
\widetilde V_1=\begin{bmatrix}
 1	&0	&0	&3	&0	&1	&2	&0	&0	&0\\
 0	&0	&0	&-1	&\frac{1}{2}	&1	&0	&0	&1	&\frac{5}{3}
\end{bmatrix}^T.
\end{equation}
If $\widetilde V$ is defined as the matrix consisting of the columns of $V$ that do not correspond to $\mathscr J_2(0)$ in addition to the columns of $\widetilde V_1$~\eqref{eq:widetildeV1}, it can be shown that $\widetilde A = \widetilde V J \widetilde V^{-1}$ does not equal $A$. Nevertheless, the oblique projection matrices~\eqref{eq:spectralprojectormatrix} corresponding to $\widetilde V_1$~\eqref{eq:widetildeV1} and~$V_3$~\eqref{eq:ex:V3} onto the  Jordan subspaces are identical; that is, the GFT~\eqref{eq:gft} is equivalent for both eigenvector matrices, and graphs~$\mathcal G(A)$ and $\mathcal G(\widetilde A)$ are in the same Jordan equivalence class corresponding to $\mathscr J_2(0)=\mathrm{span}\{V_3\}$. The total variation of~$\widetilde V_1$ with respect to $\widetilde A$ is
\begin{equation}
\mathrm{TV}_{G}(\widetilde V_1) = \left\| \widetilde V_1 - \widetilde A \widetilde V_1\right\|_1 = 1.452.
\end{equation}
Thus,~$\widetilde V_1$ does not achieve the class total variation~\eqref{eq:ex:classtotalvar}.

\section{Limitations}
\label{sec:limitations}
The Jordan equivalence classes discussed in Section~\ref{sec:jordanequiv} show that there are degrees of freedom over graph topologies with defective adjacency matrices that enable the GFT to be equivalent over multiple graph structures. It may be sufficient to find these classes by traversing the graph once (with total time complexity $O(\left| V\right| + \left| E\right|)$) and then determining the Jordan normal form of the underlying graph because of the acyclic and cyclic structures within the graph; see~\cite{cardon2011jordan,nina2013jordan} and more details in Section~\ref{sec:jordanequiv}.

On the other hand, not all graphs have structures that readily reveal their Jordan equivalence classes. For example, arbitrary directed, sparse matrices such as road networks or social networks may have complex substructures that require a full eigendecomposition before determining the corresponding Jordan equivalence class. Inexact eigendecomposition methods are useful to approximate the GFT in this case. In particular, the authors explore such a method in~\cite{deriAIM2016}.

%
%
%
%
\section{Conclusion}
\label{sec:conclusion}
This paper characterizes two equivalence classes of graph structures that arise from the spectral projector-based GFT formulation of~\cite{deriGFT2016}. Firstly, isomorphic equivalence classes ensure that the GFT is equivalent with respect to a given node ordering. This allows the exploitation of banded matrix structures that permit efficient eigendecomposition methods. Secondly, Jordan equivalence classes show that the GFT can be identical over graphs of different topologies. Certain types of graphs have Jordan equivalence classes that can be determined by a single traversal over the graph structure, which means that the eigenvector matrix can potentially be chosen for a simpler matrix topology. For more general graphs for which the equivalence class cannot be easily determined, inexact methods such as those proposed in~\cite{deriAIM2016} provide a means to computing the spectral projector-based GFT.

Lastly, a total variation-based ordering of the Jordan subspaces is proposed. Since the total variation is dependent on the particular choice of Jordan basis, we propose a class variation-based ordering that is defined by the Jordan equivalence class of the graph.
\bibliographystyle{IEEEbib}
\bibliography{refs} 
\end{document}